\documentclass[a4paper,11pt]{article}
\usepackage{amsmath}
\usepackage{amsfonts}
\usepackage{amssymb}
\usepackage{graphicx,epsfig}
\usepackage{hyperref}
\numberwithin{equation}{section}
\usepackage[svgnames]{xcolor}
\usepackage{cite}
\usepackage{array}
\usepackage{multirow}
\usepackage{rotating}
\usepackage{lscape}
\makeatletter
\renewcommand*\env@matrix[1][\arraystretch]{%
	\edef\arraystretch{#1}%
	\hskip -\arraycolsep
	\let\@ifnextchar\new@ifnextchar
	\array{*\c@MaxMatrixCols c}}
\makeatother 

\begin{document}
	\title {Classes of exact solutions for the massless Dirac particle in the $C$-metric}
	\author{Priyasri Kar\\[0.2cm]
		{\it Department of Physical Sciences,}\\ {\it Indian Institute of Science Education and Research Kolkata, Mohanpur 741246, India}\\[0.2cm]
		{\tt pk12rs063@iiserkol.ac.in}
	}
	\maketitle

	\begin{abstract}
		The massless Dirac particle in the $C$-metric, representing the exterior gravitational field of a uniformly accelerating black hole, is studied. Classes of (quasi-)polynomial solutions to the radial and the polar parts of the Dirac equation, each of which is equivalent to the general Heun equation~(GHE), are obtained exploiting the underlying $su(1,1)$ algebraic structures of the GHE.
	\end{abstract}
	
	\begin{center}
	\end{center}
	
	\section{Introduction}\label{sec_intro}
	
	The $C$-metric is an exact solution of Einstein's equations, that represents the exterior gravitational field of a uniformly accelerating Schwarzschild black hole. It was first found by Levi-civita~\cite{levicivita} and its mathematical properties were the subject of many investigations~\cite{kramer_et_al}. The $C$-metric got its place in a famous classification of exact space-times~\cite{ehlers_kundt} and few years later came its physical interpretation~\cite{kinnerslay_walker_1970,bonnor,cornish_uttley}. A comprehensive account of the historical development of knowledge in this field and the mathematical and physical properties of the metric may be found in Ref.~\cite{exact_spacetimes_book}.
	
	The present focus is the dynamics of a massless spin-$1/2$ particle in the $C$-metric, which was studied in Ref.~\cite{bini_bit_ger}. The radial and the polar parts of the Dirac equation were found to be equivalent, up to some transformations, to the General Heun Equation (GHE), whereas the equation has standard dependence on the azimuthal and temporal variables due to the existence of the corresponding Killing vectors. Hence, by finding solutions to the GHEs corresponding to the radial and polar part, it is possible to have a comprehensive idea about the behavior of the Dirac spinor. The GHE, a Fuchsian equation with four regular singularities at $0, \ 1, \ a(\ne 0,1) \mbox{ and} \ \infty$, is a natural generalization of the famous hypergeometric equation (three regular singularities at $0, \ 1 \mbox{ and} \ \infty$). In the canonical form, it reads~\cite{ronveauxbook}
	\begin{equation}\label{heuneqn}
	\frac{d^2y}{dz^2}+\left(\frac{\gamma}{z}+\frac{\delta}{z-1}+\frac{\varepsilon}{z-a}\right)\frac{dy}{dz}+\frac{\alpha\beta
		z-q}{z(z-1)(z-a)}y(z)=0,
	\end{equation}
	where, the exponents at these singularities are $(0, \ 1-\gamma)$, $(0,\ 1-\delta)$, 
	$(0, \ 1-\varepsilon)$ and $(\alpha, \ \beta)$, respectively. Here $q$ plays 
	the role of the \emph{eigen parameter}. The equation, being a second order linear 
	homogeneous equation with $4$ regular singularities, satisfies the Fuchsian condition: 
	\begin{equation}\label{fuchsiancondn}
	\gamma+\delta+\varepsilon=\alpha+\beta+1,
	\end{equation}
	allowing elimination of $\varepsilon$ in favor of the others. The General Heun Equation and its confluent versions (together known as the Heun class of equations) are too important for their frequent appearance in different branches of theoretical physics~\cite{ronveauxbook,slavyanov_lay,hortacsu_main,heunprjct}, among which, the field of general relativity is worth special mention~\cite{hortacsu_equivalent_in_astro,fiziev_teukolsky,fiziev_staicova_2,vieira_bezerra,vieira_bezerra_2,batic_schmid_2,birkandan_cvetic,schzschld_heun,kerr_heun}. Yet, the process of finding analytical solutions (both local and global) to them is not straightforward. Most of the widely used methods for finding solutions to linear second-order differential equations do not quite work for them. For example, unlike the hypergeometric equation which was factorized by Schr\"{o}dinger long back, the Heun class has not yet been factorized for general values of parameters~(some progress have been achieved along these lines but with parametric constraints). Moreover, their Frobenius series type local solutions involve three-term recursion relations, for which analytical closed-form solutions are not available as yet.
	
	Despite all these difficulties, there is one property of the Heun class, that aids the process of finding solutions to a considerable extent. The Heun class of equations are known to be quasi-exactly solvable~\cite{turbiner,turbiner_big,turbiner_qes_introduced,shif_tur,ush_book,lopez_kamran,brihaye_qes_heun,pauli_eqn}, which means that the corresponding Heun operator preserves a finite number of polynomial or quasi-polynomial type global solutions (too important from both physical and mathematical points of view, due to their compact form and validity throughout the entire complex plain with possible exceptions at the singular points, that can be avoided using proper branch cuts) under certain parametric conditions~\cite{turbiner,turbiner_big,turbiner_heun_hamiltonian,degree_half,my_2nd_paper}. A mathematically equivalent statement of the above is that the GHE can be cast in terms of differential operators satisfying $su(1,1)$ algebra in the monomial basis and that some finite-dimensional representation of $su(1,1)$ serves as the solution space of the GHE~\cite{turbiner_big}. There exist different sets of $su(1,1)$ generators in terms of which the Heun class of equations can be cast. Among them, the following set, proposed for the first time by Sophus Lie,
	\begin{equation}\label{lie-generators}
	\mathcal{J}^+= z^2\dfrac{d}{dz}-2j z, \quad
	\mathcal{J}^0= z\dfrac{d}{dz}-j \quad \mbox{and} \quad 
	\mathcal{J}^-= \dfrac{d}{dz},
	\end{equation}
	has been used in the Refs.~\cite{turbiner,turbiner_big,turbiner_heun_hamiltonian}. These generators, of \emph{degrees}\footnote{The degree $d$, of an operator $O_d$, is defined as the change in the power of a monomial $\{z^c: c \in \mathbb{C}\}$, when acted upon by it, i.e., $O_d \ z^c \propto z^{c+d}$. } $\pm 1$ and $0$ and involving a single parameter $j$, yields all the polynomial solutions admitted by the equation. In Ref.~\cite{degree_half}, a set of  \emph{bi-parametric} generators, of degrees $\pm 1/2$ and $0$, has been used to cast the GHE and some additional quasi-polynomial solutions of the terminated Frobenius series form:
	\begin{equation}\label{quasi_poly}
	y(z)=z^{c}P_N(z); \quad \{c \in \mathbb{C}, \ P_N(z): \mbox{polynomial in } z\},
	\end{equation}
	were found as a result. The algebraization, however, requires the GHE to satisfy a parametric condition, which is, the regular singularities at $0$ and $\infty$ of the GHE must be \emph{elementary}\footnote{Any regular singularity can be characterized by two exponents $\rho_1$ and $\rho_2$, which are the two roots of the indicial equation. If $|\rho_1-\rho_2|=1/2$, the corresponding regular singularity is known as an elementary singularity. They bear some special significance in the sense that all regular and irregular singularities can be obtained by the coalescence of two and three or more elementary singularities, respectively~\cite{ince}.}. Then in Ref.~\cite{my_2nd_paper}, a set of \emph{bi-parametric} generators of degrees $\pm 1$ and $0$ (like the Lie generators), have been used, which made the algebraization of the GHE (and the whole Heun class) possible without any algebraic constraint. It has also been shown that this method can find any (quasi-)polynomial solution of the form~(\ref{quasi_poly}), which of course includes all the polynomial solutions. Among all these solutions, the ones for which $c \neq 0$, are \emph{additional solutions}~\cite{my_2nd_paper} in the sense that they can not directly be obtained using the Lie generators~(\ref{lie-generators}). Thus in terms of algebraizable equations and also obtainable solutions, this last set of $su(1,1)$ generators, of integer degrees and involving two parameters, gives the maximum coverage.
	
	In this paper, this set of generators is be used to cast the GHEs corresponding to the radial and the polar parts of the Dirac equation, and both the equations are found to admit classes of countable infinities of (quasi-)polynomial solutions (known as Heun polynomials in case of GHE). The organization of the paper is as follows: in section~\ref{sec_rad_pol_&_ghe}, the two GHEs are obtained from the radial and polar parts of the Dirac equation for the massless spin-$1/2$ particle in the $C$-metric. Section~\ref{sec_algebra_&exct_solns} introduces the $su(1,1)$ generators and the proposed form to cast the Heun operator in terms of them. Fundamentals of obtaining Heun polynomials exploiting the $su(1,1)$ representation theory are also discussed. In section~\ref{sec_polar}, it is shown that the equation admits two countably infinite sets of quasi-polynomial solutions to the polar part and in section~\ref{sec_radial} it is shown that the equation admits one countable infinity of polynomials and one countable infinity of quasi-polynomial solutions to the radial part. The sets of quasi-polynomial solutions are additional solutions. The first few solutions of all these four sets are listed. Finally, in section~\ref{sec_conclusion}, the results are summarized and concluded.
	
	\section{The Dirac equation: the polar and the radial parts}\label{sec_rad_pol_&_ghe}
	
	The line element of the $C$-metric, representing the exterior gravitational field of a uniformly accelerating black hole, in spherical-like co-ordinates is given by~\cite{exact_spacetimes_book}
	\begin{subequations}\label{C-metric}
		\begin{align}
		ds^2=&\dfrac{1}{\Omega^2(r,\theta)}\left(\dfrac{Q(r)}{r^2}dt^2-\dfrac{r^2}{Q(r)}dr^2-\dfrac{r^2}{P(\theta)}d \theta^2-r^2P(\theta) \mbox{sin}^2 \ \theta d \phi^2 \right), \label{metric}\\
		\nonumber \mbox{where},\\
		Q(r)=&r\left(r-2M\right)\left(1-A^2r^2\right), \quad P(\theta)=1-2MA \ \mbox{cos}\theta, \quad \Omega(r,\theta)=1-Ar \ \mbox{cos} \theta \label{metric_functions}
		\end{align}
	\end{subequations}
	and the constants $M(\ge0)$ and $A(\ge0)$ are the mass and acceleration of the black hole, respectively. This metric reduces to two well-known metrics in two different parametric limits: the spherically symmetric Schwarzschild metric for $A=0$ and the Rindler metric representing flat spacetime with uniform acceleration for $M=0$. The properties of the metric are as follows: The physical region of interest for the solution of the radial equation lies between the Schwarzschild horizon $r=2M$ and the Rindler horizon $r=1/A$, i.e., one must have $2M<r<1/A$. This requires $2M<1/A\ (\mbox{or}\ MA<1/2)$, which ensures that the metric signature is preserved ($P(\theta)>0$) for all $\theta \in \left[0, \pi \right]$. In other words, the acceleration parameter $\eta (\equiv 2MA)$ of the black hole is less than 1 in the physical region of interest. Moreover, if one is to restrict oneself within the realm of $C$-metric (excluding the pure Schwarzschild $(A=0)$ or pure Rindler $(M=0)$ subcases), which is the focus of this chapter, then only positive definite values of $\eta$ should be considered. Hence, for the present purpose, the allowed range of $\eta$ will be
	\begin{equation}\label{etarange}
	0<\eta<1.
	\end{equation}
	There is a conical singularity at $\theta=0$, which can be avoided with the choice of range of $\phi$ given by $\phi \in \left[0,2 \pi/P(0)\right]$.
	
	The Dirac equation for the massless spin 1/2 particle in the $C$-metric is obtained~\cite{bini_bit_ger} using Newman-Penrose formalism in Kinnersley type null frame~\cite{kinnersley}. Following the notation of Ref.~\cite{bini_bit_ger}, the 4-component Dirac spinor $\psi$ is given by
	\begin{equation}
	\psi^T=e^{-\iota(\omega t-m \phi)}\begin{pmatrix} 
	F_1, \ \
	F_2, \ \
	-G_2,\ \
	G_1
	\end{pmatrix}
	\end{equation}
	where,
	\begin{subequations}
		\begin{align}
		F_1&=\dfrac{1}{\sqrt{2}r} \ \dfrac{\Omega}{P(\theta)^{1/4}}  \ R_{-1/2}\left(r\right)  \ S_{-1/2}\left(\theta\right) \qquad  
		F_2=\dfrac{\Omega}{\sqrt{Q(r)}} \ \dfrac{\Omega}{P(\theta)^{1/4}} \ R_{1/2}\left(r\right)  \ S_{1/2}\left(\theta\right)\\
		G_1&=\dfrac{\Omega}{\sqrt{Q(r)}} \ \dfrac{\Omega}{P(\theta)^{1/4}} \ R_{1/2}\left(r\right)  \ S_{-1/2}\left(\theta\right) \quad \hspace{.1cm} 
		G_2=\dfrac{1}{\sqrt{2}r} \  \dfrac{\Omega}{P(\theta)^{1/4}} \ R_{-1/2}\left(r\right)  \ S_{1/2}\left(\theta\right).
		\end{align}
	\end{subequations}
	The exponential form of the azimuthal and the temporal parts of the spinor is due to the existence of the time-like and the rotational Killing vectors $\partial_t$ and $\partial_\phi$ in the $C$-metric. The polar functions $S_{\pm 1/2}\left(\theta\right)$ and the radial functions $R_{\pm 1/2}\left(r\right)$ separately satisfy two pairs of first-order coupled differential equations, which lead to one second-order differential equation for each of the four radial and polar functions.
	
	\subsection{The general Heun equation obtained from the polar equation}
	
	The second order equations for the polar functions $S_{\pm 1/2}$ read
	\begin{eqnarray}\label{Dirac_polar}
	\nonumber \dfrac{d^2S_{\pm 1/2}}{d\theta^2}+\left(\dfrac{P_{\theta}}{2P}+\cot\theta\right)\dfrac{dS_{\pm 1/2}}{d\theta} \hspace{4.85cm} \\
	-\dfrac{1}{2} \left[1-\dfrac{2\lambda^2}{P} -\dfrac{P_{\theta}\left(P\cos\theta \mp 2m\right)}{2P^2\sin\theta}+\dfrac{\left(P\cos\theta \pm 2m\right)}{2P^2\sin^2\theta}\right]S_{\pm 1/2}(\theta)=0
	\end{eqnarray}
	where, $P_{\theta}=\partial P/\partial \theta=\eta \sin \theta$. Effecting the F-homotopic transformation
	\begin{equation}\label{polar_homotopy}
	S_{\pm 1/2}(\theta)=z^{k_0}(z-1)^{k_1}\left(\dfrac{2\eta z-\eta-1}{2\eta}\right)^{k_a}y(z)
	\end{equation}
	in terms of the new variable $z=\cos^2(\theta/2)$, where
	\begin{eqnarray}\label{polar_expnts}
	k_0=\dfrac{\eta\mp 2m+1}{4(\eta+1)},\quad k_1=\dfrac{\eta\mp 2m-1}{4(\eta-1)} \quad \mbox{and} \quad k_a=\dfrac{1}{4}+\dfrac{\eta^2 \pm 4m\eta-1}{4(\eta^2-1)},
	\end{eqnarray}
	one obtains
	\begin{eqnarray}\label{polar_GHE}
	\nonumber \dfrac{d^2y_{\pm}}{dz^2}+\left[\dfrac{3\eta \mp 2m+3}{2z(\eta+1)}+\dfrac{3\eta \mp 2m-3}{2(z-1)(\eta-1)}+\dfrac{2\eta\left(3\eta^2\pm 4m\eta-3\right)}{2(2\eta z-\eta-1)(\eta^2-1)}\right]\dfrac{dy_{\pm}}{dz}&\\
	+\dfrac{6\eta z+\lambda^2-3\eta-1}{2\eta z(z-1)(2\eta z-\eta-1)}y_{\pm}(\theta)&=0,
	\end{eqnarray}
	which is a GHE of form~(\ref{heuneqn}) with parameters
	\begin{eqnarray}\label{polar_GHE_params}
	\nonumber \gamma_{\theta}=\dfrac{3\eta \mp 2m+3}{2(\eta+1)}, \quad \delta_{\theta}=\dfrac{3\eta \mp 2m-3}{2(\eta-1)}, \quad \varepsilon_{\theta}=\dfrac{\left(3\eta^2\pm 4m\eta-3\right)}{2(\eta^2-1)},\\
	\alpha_{\theta}=2, \quad \beta_{\theta}=\dfrac{3}{2}, \quad a_{\theta}=\dfrac{\eta+1}{2\eta} \quad \mbox{and} \quad q_{\theta}=-\dfrac{\lambda^2-3\eta-1}{2\eta}.
	\end{eqnarray}
	
	\subsection{The general Heun equation obtained from the radial equation}
	
	The second order equations for the radial functions $R_{\pm 1/2}$ are given by
	\begin{equation}\label{Dirac_radial}
	\dfrac{d^2R_{\pm 1/2}}{dr^2}+\dfrac{Q_r}{2Q}\dfrac{dR_{\pm 1/2}}{dr}-\dfrac{1}{Q}\left[\lambda^2-\dfrac{\omega^2r^4}{Q}\mp2\iota\omega r\left(1-\dfrac{rQ_r}{4Q}\right)\right]R_{\pm 1/2}(r)=0,
	\end{equation}
	where, $Q_r=\partial Q/\partial r$. Performing the Mobius transformation
	\begin{eqnarray}\label{radial_Mobius}
	u=\dfrac{Ar(1-\eta)}{\eta(1-Ar)},
	\end{eqnarray}
	followed by the F-homotopic transformation
	\begin{equation}\label{radial_homotopy}
	R_{\pm 1/2}(u)=z^{l_0}(z-1)^{l_1}\left(\dfrac{2\eta z-\eta+1}{2\eta}\right)^{l_a}g_{\pm}(u),
	\end{equation}
	where,
	\begin{equation}\label{radial_expnts}
	l_0=0, \quad l_1=\mp\dfrac{2\iota M \omega}{1-\eta^2}, \quad \mbox{and} \quad l_a=\mp \dfrac{\iota M \omega}{\eta(1+\eta)},
	\end{equation}
	one obtains
	\begin{equation}\label{radial_GHE}
	\dfrac{d^2g_{\pm}}{du^2}+\left[\dfrac{1}{2u}+\dfrac{\eta^2-1\pm 8\iota M \omega}{2(u-1)(\eta^2-1)}+\dfrac{\eta^2+\eta\mp 4\iota M \omega}{(2\eta u-\eta+1)(\eta+1)}\right]\dfrac{dg_{\pm}}{du}+\dfrac{\lambda^2}{u(u-1)(2\eta u-\eta+1)}g_{\pm}(u)=0
	\end{equation}
	which is a GHE of form~(\ref{heuneqn}) with parameters
	\begin{eqnarray}\label{radial_GHE_params}
	\nonumber \gamma_r=\dfrac{1}{2}, \quad \delta_r=\dfrac{\eta^2-1\pm 8\iota M \omega}{2(\eta^2-1)}, \quad \varepsilon_r=\dfrac{\eta^2+\eta\mp 4\iota M \omega}{2\eta(\eta+1)}\\
	\alpha_r=0, \quad \beta_r=\dfrac{\eta^2-\eta\pm 4\iota M \omega}{2\eta(\eta-1)}, \quad a_r=\dfrac{\eta-1}{2\eta} \quad \mbox{and} \quad q_r=\dfrac{\lambda^2}{2\eta}.
	\end{eqnarray}
	
	Thus it is found that the polar and the radial parts of the Dirac equation are equivalent to the general Heun equation up to some transformations. In the next section, the Lie algebraic structure of the GHE is exploited to obtain polynomial and quasi-polynomial type global solutions to the polar and the radial parts.
	
	\section{Lie algebraic structure of the GHE and (quasi-)polynomial solutions}\label{sec_algebra_&exct_solns}
	
	As has been discussed earlier, the present aim is to cast the GHEs coming from the polar and the radial parts of the Dirac equation in terms of a set of bi-parametric differential operators of degrees $\pm 1$ and $0$ \cite{my_2nd_paper}, satisfying $su(1,1)$ algebra in the monomial basis. Depending on the values of the equation parameters, different representation spaces of $su(1,1)$ serve as the solution spaces of the differential equation in concern. When a finite-dimensional representation is available as a solution space, linearly independent Heun polynomials, equal in number to the dimensionality of the representation space, are available as solutions to the concerned GHE. The integer degree bi-parametric $su(1,1)$ generators are given by~\cite{my_2nd_paper}
	\begin{equation}\label{newgenerators}
		J^+= z^2\dfrac{d}{dz}-2\sigma z, \quad
		J^0= z\dfrac{d}{dz}-(\sigma+\tau) \quad \mbox{and} \quad 
		J^-= \dfrac{d}{dz}-\dfrac{2\tau}{z},
	\end{equation}
	which are of degrees $+1$, $0$ and $-1$ respectively. These generators satisfy $su(1,1)$ algebra in the monomial basis with the Casimir given by
	\begin{equation}\label{construct_casimir}
	C_2=-\left(\sigma-\tau\right)\left(\sigma-\tau+1\right)=-j(j+1),
	\end{equation}
	where $j=(\sigma-\tau)$. Multiplying Eq.~(\ref{heuneqn}) with $z(z-1)(z-a)$ and writing it in the operator form as
	\begin{equation}
	\mathcal{H}y(z)=0,
	\end{equation}
	what remains is to cast $\mathcal{H}$ in terms of generators\ref{newgenerators}. For the purpose, the following ansatz is made~\cite{my_2nd_paper}:
	\begin{equation}\label{heun_as_sl2}
	\mathcal{H}=\sum_{\substack{k,l=+,0,-{}\\k > l}}c_{kl}J^kJ^l +\sum_{k=+,0,-}c_kJ^k +c,
	\end{equation}
	where $c_{kl}, c_k$ and $c$ are arbitrary constants. It is observed that one can always solve for the values of the construct parameters (consisting of $c_{kl}, c_k$, $c$ and the two parameters $\sigma$ and $\tau$ associated with the generators~(\ref{newgenerators})), such that the parameters of a general Heun equation (GHE) $\gamma, \ \delta, \ \varepsilon, \ \alpha, \ \beta$ and $a$ (see Eq.~(\ref{heuneqn})) can be expressed in terms of the construct parameters, or in other words, the GHE always has the algebraic structure given by Eq.~(\ref{heun_as_sl2}). The values of the two parameters $\sigma$ and $\tau$ are of central importance, since the quantity $(\sigma-\tau)=j$ (called the \emph{representation parameter}) determines the representation(s) of $su(1,1)$ available for a particular equation~\cite{my_2nd_paper}. Both the parameters $\sigma$ and $\tau$ can assume two values. In terms of the GHE parameters they are given as
	\begin{equation}\label{GHE_sigma_tau}
	\big\{\sigma^{(1)},\sigma^{(2)}\big\}= \{-\alpha/2, -\beta/2\} \quad \mbox{and} \quad \big\{\tau^{(1)},\tau^{(2)}\big\}=\{0, (1-\gamma)/2\}.
	\end{equation}
	
	Thus there are four possible candidates for $\{\sigma,\tau\}$ pair, viz, $\{\sigma^{(1)},\tau^{(1)}\}$, $\{\sigma^{(1)},\tau^{(2)}\}$, $\{\sigma^{(2)},\tau^{(1)}\}$ and $\{\sigma^{(2)},\tau^{(2)}\}$. When for any one or more of them the representation parameter is equal to non-negative half integer, i.e., when
	\begin{equation}\label{quasi_exact_condn}
	(\sigma-\tau)=j=N/2, \ \mbox{for some} \ N \in \mathbb{N}_0;
	\end{equation}
	the finite $(2j+1)=(N+1)$ dimensional representation of $su(1,1)$ is available as solutions space, or in other words, the corresponding general Heun operator has a finite $(N+1)$ dimensional invariant subspace in the monomial basis. In that case, one obtains $(N+1)$ linearly independent Heun polynomial solutions of the form
	\begin{equation}\label{heun_polynomial}
	y(z)=z^{2\tau}P_N(z)=z^{2\tau}\sum_{i=0}^{N}k_iz^i
	\end{equation}
	for $(N+1)$ values of the eigenparameter $q$ (Eq.~(\ref{heuneqn})), where $P_N(z)$ is polynomial in $z$ of order $N=2j$. The coefficients $k_i$ and the eigenvalues $q$ are given (Appendix~\ref{appendix_matrix_forms}) by the eigenvectors and the eigenvalues of the matrix~(\ref{GHE_tridiag_poly}), discussed in Ref.~\cite{ciftci_hall_saad_dogu}, when $\tau=0$ and by those of matrix~(\ref{GHE_tridiag_quasi}), discussed in Ref.~\cite{my_2nd_paper}, when $\tau \neq 0$. With this knowledge of the algebraic properties of the GHE, the next job is to look for (quasi-)polynomial type global solutions to the polar and the radial parts.
	 
	\section{Heun polynomials solutions to the polar part}\label{sec_polar}
	 
	From Eqs.~(\ref{GHE_sigma_tau}) and (\ref{polar_GHE_params}), the $\sigma$ and $\tau$ values for the polar GHE~(\ref{polar_GHE}) are found to be
	\begin{equation}\label{polar_sigma_tau}
	\left\{\sigma^{(1)}_{\theta},\sigma^{(2)}_{\theta}\right\}= \left\{-1, -{3\over 4}\right\} \quad \mbox{and} \quad \left\{\tau^{(1)}_{\theta},\tau^{(2)}_{\theta}\right\}=\left\{0, -{(\eta\mp 2m+1)\over 4(\eta+1)}\right\}.
	\end{equation}
	
	Among the four possible $\{\sigma_{\theta},\tau_{\theta}\}$ pairs it is easy to see that $\big\{\sigma^{(1)}_{\theta},\tau^{(1)}_{\theta}\big\}$ and $\big\{\sigma^{(2)}_{\theta},\tau^{(1)}_{\theta}\big\}$ do not satisfy condition~(\ref{quasi_exact_condn}). For the other two pairs $\big\{\sigma^{(1)}_{\theta},\tau^{(2)}_{\theta}\big\}$ and $\big\{\sigma^{(2)}_{\theta},\tau^{(2)}_{\theta}\big\}$ one can look for parametric conditions leading to condition~(\ref{quasi_exact_condn}).
	
	\subsection{Conditions for Heun polynomials from the $\left\{\sigma^{(1)}_{\theta},\tau^{(2)}_{\theta}\right\}$}
	
	The demand that the pair $\big\{\sigma^{(1)}_{\theta},\tau^{(2)}_{\theta}\big\}$ must satisfy condition~(\ref{quasi_exact_condn}), i.e., $\big(\sigma^{(1)}_{\theta}-\tau^{(2)}_{\theta}\big)=N/2$ for some $N \in \mathbb{N}_0$ requires
	\begin{equation}\label{eta_m_reln_1}
	\eta=-\dfrac{2N\pm 2m+3}{2N+3} \quad \mbox{or} \quad m=\mp\dfrac{\left(\eta+1\right)\left(2N+3\right)}{2}
	\end{equation}
	for $S_{\pm 1/2}$. Now imposing the physically allowed range of $\eta$~(Eq.~(\ref{etarange})) on Eq.~(\ref{eta_m_reln_1}) one obtains the following conditions on $m$:
	\begin{equation}\label{allowed_m_values_1}
	-(2N+3)<m<-\dfrac{2N+3}{2} \quad \mbox{for} \quad S_{+1/2} \quad \mbox{and} \quad \dfrac{2N+3}{2}<m<(2N+3) \quad \mbox{for} \quad S_{-1/2}
	\end{equation}
	Now, an important point to notice here is that the two equations~(\ref{Dirac_polar}) for the two polar functions $S_{\pm 1/2}$ or the corresponding GHEs~(\ref{polar_GHE}) differ only in the signature of the azimuthal quantum number $m$. Hence, the symmetry in the allowed values of $m$ for the two polar functions (which are of exactly opposite sign~(\ref{allowed_m_values_1})), implies that putting the allowed $m$ values back into the two equations will make them identical. Consequently, the solutions to the two polar equations will also be identical (modulo for interpretation). Next one has to find out, for a particular $N$, the $m$ values leading to $\big\{\sigma^{(1)}_{\theta},\tau^{(2)}_{\theta}\big\}$ pair that satisfies the quasi-exact solvability condition~(\ref{quasi_exact_condn}). For convenience it is worked out for $S_{-1/2}$, the allowed values for $S_{+1/2}$ will evidently be the opposite due to the symmetry discussed above. Keeping in mind that the $m$ values must be half-(odd)integers, the lowest allowed $m$ value~(from Eq.~(\ref{allowed_m_values_1})) for $S_{-1/2}$ is $(2N+5)/2$ and the highest $m$ value is $2N+5/2$. The allowed $m$ values are evidently the half-(odd)integers between the highest and lowest $m$ values (obviously including both of them), the total number counting to $(N+1)$. Thus for $\forall N\in\mathbb{N}_0$, there exists $(N+1)$ values for $m$~(given by $\{\mp(2N+5)/2,\mp(2N+7)/2,\dots,\mp(4N+5)/2\}$), that satisfy the condition~(\ref{quasi_exact_condn}) for $S_{\pm 1/2}$. The $\eta$ values corresponding to the allowed $m$ values can be obtained from Eq.~(\ref{eta_m_reln_1}). The allowed $m$ and $\eta$ values for the first few $N$ are listed in Table~\ref{tab_eta_m_1}, where the $m$ values with the upper sign are the allowed values for $S_{+1/2}$ and those with the lower sign are the allowed values for $S_{-1/2}$.
	\begin{table}[!ht]
		\caption{\label{tab_eta_m_1}{Allowed $m$ values ($m_{\pm}$ for $S_{\pm1/2}$) and $\eta$ values corresponding to different $su(1,1)$ representations for the $\left\{\sigma^{(1)}_{\theta},\tau^{(2)}_{\theta}\right\}$ case}}
		\centering
		\begin{tabular}{|c|c|c|c@{}|}
			\hline
			\textbf{$N$ value} & \textbf{Dim of rep} & \textbf{$m_{\pm}$} & \textbf{$\eta$} \\
			\hline
			$0$ & $1$(Singlet) & $\mp5/2$ & $2/3$ \\ 
			\hline
			$1$ & $2$(Doublet) & \begin{tabular}{@{}c@{}}$\mp7/2$ \\ $ \mp9/2$ \end{tabular} & \begin{tabular}{@{}c@{}}$2/5$ \\ $ 4/5$ \end{tabular}\\
			\hline 
			$2$ & $3$(Triplet) & \begin{tabular}{@{}c@{}c@{}}$\mp9/2$ \\ $ \mp11/2$ \\ $\mp 13/2$ \end{tabular} & \begin{tabular}{@{}c@{}c@{}}$2/7$ \\ $ 4/7$ \\ $6/7$ \end{tabular}\\
			\hline
			$\vdots$ & $\vdots$ & $\vdots$ & $\vdots$\\
			\hline 
			$N$ & \begin{tabular}{c}
				$N+1$ \\ ($(N+1)$-plet)
			\end{tabular} & \begin{tabular}{@{}c@{}c@{}c@{}}$\mp(2N+5)/2$ \\ $ \mp(2N+7)/2$ \\ $\vdots$ \\ $\mp(4N+5)/2$ \end{tabular} & \begin{tabular}{@{}c@{}c@{}c@{}}$2/(2N+3)$ \\ $ 4/(2N+3)$ \\ $\vdots$ \\ $(2N+2)/(2N+3)$ \end{tabular} \rule{0pt}{2.9ex} \\
			\hline
		\end{tabular}
	\end{table}
	Thus $\forall N \in \mathbb{N}_0$ there exists $(N+1)$ allowed $\{m,\eta\}$ pair, for each of which the quasi-exact solvability condition~(\ref{quasi_exact_condn}) is satisfied, i.e. (section~\ref{sec_algebra_&exct_solns}), for each of which $(N+1)$ dimensional representation space of $su(1,1)$ serve as the solution space of the GHE~(\ref{polar_GHE}). Hence for each allowed $\{m,\eta\}$ pair, $(N+1)$ linearly independent solutions of the form~(\ref{heun_polynomial}) are obtained (section~\ref{sec_algebra_&exct_solns}) to the GHE~(\ref{polar_GHE}), corresponding to $(N+1)$ values of the eigen-parameter $q_{\theta}$~(Eq.~\ref{polar_GHE_params}). Evidently $\forall N \in \mathbb{N}_0$ there exists a total of $(N+1)^2$ solutions of the form~(\ref{heun_polynomial}) to the GHE~(\ref{polar_GHE}). Since this is true $\forall N \in \mathbb{N}_0$, hence every finite dimensional representation of $su(1,1)$ serve as solution space for the GHE~(\ref{polar_GHE}). The coefficients $k_i$ in the solutions~(\ref{heun_polynomial}) and the corresponding eigenvalues $q_{\theta}$~(Eq.~\ref{polar_GHE_params}) are given by the eigenvectors and the eigenvalues of the matrix~(\ref{GHE_tridiag_quasi}), which is the matrix form of a General Heun operator satisfying condition~(\ref{quasi_exact_condn}) for non-zero $\tau$ (here, $\tau^{(2)}_{\theta}$). These are \emph{additional} Heun polynomials, obtained as a result of casting the Heun equation in terms of generators~(\ref{newgenerators}). Plugging these GHE solutions $y_{\pm}(z)$ into Eq.~(\ref{polar_homotopy}) one obtaines a countable infinity of exact global solutions to the polar part~(\ref{Dirac_polar}) of the Dirac equation and the corresponding values of the separation constant ($\lambda$) are found using $q_{\theta}-\lambda$ relation given by Eq.~(\ref{polar_GHE_params}). Solutions upto triplet are listed in Table~\ref{tab_polar_sols_1}. Among them, analytical expressions for eigenvalues and eigenvectors have been provided for the singlet and the doublet. Analytical expressions from triplet onwards become huge and cumbersome. The eigenvalues $q_{\theta}$ for triplet are given by
	\begin{eqnarray}
	\nonumber q_{{\theta}_{1,2,3}}=\dfrac{1}{2\eta} \times \mbox{Roots}\left[x^3+\left(11+33\eta\right)x^2+\left(24+186\eta+258\eta^2\right)x \right.\\
	\left.+96\eta+384\eta^2+288\eta^3==0,x\right]
	\end{eqnarray}
	The corresponding separation constants will be given by $\displaystyle \lambda^2_{{\theta}_{1,2,3}}=-2\eta q_{{\theta}_{1,2,3}}+3\eta+1$ (Eq.~(\ref{polar_GHE_params})). Similar expressions are obtained for the eigenvectors (or, the coefficient $k_i$ of the Heun polynomials) as well. For brevity of presentation numerical values have been provided for eigenvalues and eigenvectors, corresponding to the triplet. For example, the values of the above constants $\lambda^2_{{\theta}_{1,2,3}}$ for $\eta=2/7$ (the first allowed $\eta$ value for triplet in Table~\ref{tab_eta_m_1}), rounded off to four decimal places, are obtained as
	\begin{equation}
	\lambda^2_{{\theta}_{1,2,3}}=15.3834, \ 7.9669, \ 2.6496,
	\end{equation}
	as have been listed in Table~\ref{tab_polar_sols_1}. Same method has been repeated for the other two values of $\eta$ leading to triplet solutions.
	
	{
	\renewcommand{\arraystretch}{1.3}
	\begin{table}[!ht]
		\caption{\label{tab_polar_sols_1}{First few solutions corresponding to $\left\{\sigma^{(1)}_{\theta},\tau^{(2)}_{\theta}\right\}$ pair for both polar functions $S_{\pm 1/2}$}}
		\centering
		\small
		\begin{tabular}{|@{}c@{}|c|@{}c@{}|@{}c@{}|}\hline
			\begin{tabular}{c}
				$su(1,1)$ \\ \textbf{Rep}
			\end{tabular} & $\eta $ & \textbf{$\lambda^2$} & \textbf{Solutions \big($S_{\pm 1/2}$\big)} \\
			\hline
			Singlet & $2/3$ & $3$ & $z^{-2}(z-1)^{-7/2}(z-5/4)^{7/2}$ \\
			\hline
			\multirow{4}{*}{Doublet} & \multirow{2}{*}{$2/5$} & $\frac{1}{10} \left(55+\sqrt{753}\right)$ & $z^{3/2}(z-1)^{-8/3}(z-7/4)^{13/6}\left[\frac{1}{28} \left(-33+\sqrt{753}\right)z^{-2}+z^{-3}\right]$\\
			&  & $\frac{1}{10} \left(55-\sqrt{753}\right)$ & $z^{3/2}(z-1)^{-8/3}(z-7/4)^{13/6}\left[\frac{1}{28} \left(-33-\sqrt{753}\right)z^{-2}+z^{-3}\right]$ \\ \cline{2-4}
			& \multirow{2}{*}{$4/5$} & $\frac{1}{10} \left(85+3\sqrt{193}\right)$ &  $z^{3/2}(z-1)^{-11}(z-9/8)^{21/2}\left[\frac{1}{12} \left(-17+\sqrt{193}\right)z^{-2}+z^{-3}\right]$\\
			&  & $\frac{1}{10} \left(85-3\sqrt{193}\right)$ &  $z^{3/2}(z-1)^{-11}(z-9/8)^{21/2}\left[\frac{1}{12} \left(-17-\sqrt{193}\right)z^{-2}+z^{-3}\right]$\\\hline
			\multirow{9}{*}{Triplet} & \multirow{3}{*}{$2/7$} & $15.3834$ & $z^2(z-1)^{-29/10}(z-9/4)^{19/10}\left[0.0207z^{-2}-0.3261z^{-3}+0.9451z^{-4}\right]$\\
			&  & $7.9669$ & $z^2(z-1)^{-29/10}(z-9/4)^{19/10}\left[-0.1273z^{-2}+0.9075z^{-3}-0.4002z^{-4}\right]$\\
			&  & $2.6496$ & $z^2(z-1)^{-29/10}(z-9/4)^{19/10}\left[0.7219z^{-2}-0.6674z^{-3}+0.1830z^{-4}\right]$\\ \cline{2-4}
			& \multirow{3}{*}{$4/7$} & $22.1558 $ & $z^2(z-1)^{-37/6}(z-11/8)^{31/6}\left[0.0383z^{-2}-0.4339z^{-3}+0.9001z^{-4}\right]$\\
			&  & $11.8062$ & $z^2(z-1)^{-37/6}(z-11/8)^{31/6}\left[-0.1739z^{-2}+0.9225z^{-3}-0.3445z^{-4}\right]$\\
			&  & $4.0380$ & $z^2(z-1)^{-37/6}(z-11/8)^{31/6}\left[0.7837z^{-2}-0.6052z^{-3}+0.1399z^{-4}\right]$\\ \cline{2-4}
			& \multirow{3}{*}{$6/7$} & $29.0682$ & $z^2(z-1)^{-45/2}(z-13/12)^{43/2}\left[0.0487z^{-2}-0.4826z^{-3}+0.8745z^{-4}\right]$\\
			&  & $15.5775$ & $z^2(z-1)^{-45/2}(z-13/12)^{43/2}\left[-0.1989z^{-2}+0.9289z^{-3}-0.3124z^{-4}\right]$\\
			&  & $5.3543$ & $z^2(z-1)^{-45/2}(z-13/12)^{43/2}\left[0.8161z^{-2}-0.5658z^{-3}+0.1177z^{-4}\right]$\\ \hline
		\end{tabular}
	\end{table}
}
	
	\subsection{Conditions for Heun polynomials from the $\left\{\sigma^{(2)}_{\theta},\tau^{(2)}_{\theta}\right\}$}
	
	For the pair $\big\{\sigma^{(2)}_{\theta},\tau^{(2)}_{\theta}\big\}$ to satisfy condition~(\ref{quasi_exact_condn}), i.e., $\big(\sigma^{(2)}_{\theta}-\tau^{(2)}_{\theta}\big)=N/2$ for some $N \in \mathbb{N}_0$, the following relation between $\eta$ and $m$ is found to be necessary:
	\begin{equation}\label{eta_m_reln_2}
	\eta=-\dfrac{N\pm m+1}{N+1} \quad \mbox{or} \quad m=\mp\left(\eta+1\right)\left(N+1\right)
	\end{equation}
	for $S_{\pm 1/2}$. The range of $\eta$ given by Eq.~(\ref{etarange}) provides the allowed values of $m$ as in the previous case:
	\begin{equation}\label{allowed_m_values_2}
	-2(N+1)<m<-(N+1) \quad \mbox{for} \quad S_{+1/2} \quad \mbox{and} \quad (N+1)<m<2(N+1) \quad \mbox{for} \quad S_{-1/2}
	\end{equation}
	Once again, the symmetry in the allowed $\eta$ values for the two polar functions imply that on putting these allowed values back, the equations and hence, the solutions will be identical. Remembering that the allowed $m$ values must be half-(odd)integers one obtains from Eq.~(\ref{allowed_m_values_2}) the lowest and highest allowed values of $m$, which are $(N+3/2)$ and $(2N+3/2)$ for $S_{-1/2}$. The allowed values are the half-(odd)integers between (and including) these two, the total number being $(N+1)$ for this case too. Thus the allowed values of $m$ that satisfy condition~(\ref{quasi_exact_condn}) for $S_{\pm1/2}$, are $\{\mp(N+3/2),\mp(N+5/2),\dots,\mp(2N+3/2)\}$ $\forall N \in N_0$. The $\eta$ values corresponding to the allowed $m$ values can be obtained from Eq.~(\ref{eta_m_reln_2}). The allowed $\{m,\eta\}$ pairs for different $N$ are listed in Table~\ref{tab_eta_m_2}.
	\begin{table}[!ht]
		\centering
		\caption{\label{tab_eta_m_2}{Allowed $m$ values ($m_{\pm}$ for $S_{\pm1/2}$) and $\eta$ values corresponding to different $su(1,1)$ representations for the $\left\{\sigma^{(2)}_{\theta},\tau^{(2)}_{\theta}\right\}$ case}}
		\begin{tabular}{|c|c|c|c|}
			\hline
			\textbf{$N$ value} & \textbf{Dim of rep} & \textbf{$m_{\pm}$} & \textbf{$\eta$} \\
			\hline
			$0$ & $1$(Singlet) & $\mp3/2$ & $1/2$ \\ 
			\hline
			$1$ & $2$(Doublet) & \begin{tabular}{@{}c@{}}$\mp5/2$ \\ $ \mp7/2$ \end{tabular} & \begin{tabular}{@{}c@{}}$1/4$ \\ $ 3/4$ \end{tabular}\\
			\hline 
			$2$ & $3$(Triplet) & \begin{tabular}{@{}c@{}c@{}}$\mp7/2$ \\ $ \mp9/2$ \\ $\mp 11/2$ \end{tabular} & \begin{tabular}{@{}c@{}c@{}}$1/6$ \\ $ 1/2$ \\ $5/6$ \end{tabular}\\
			\hline
			$\vdots$ & $\vdots$ & $\vdots$ & $\vdots$\\
			\hline
			$N$ & \begin{tabular}{c}
				$N+1$ \\ ($(N+1)$-plet)
			\end{tabular} & \begin{tabular}{@{}c@{}c@{}c@{}}$\mp(2N+3)/2$ \\ $ \mp(2N+5)/2$ \\ $\vdots$ \\ $\mp(4N+3)/2$ \end{tabular} & \begin{tabular}{@{}c@{}c@{}c@{}}$1/2(N+1)$ \\ $ 3/2(N+1)$ \\ $\vdots$ \\ $(2N+1)/2(N+1)$ \end{tabular}\\
			\hline
		\end{tabular}
	\end{table}
	For each of the allowed $\{m,\eta\}$ pair (clearly there are $(N+1)$ of them $\forall N \in N_0$), $(N+1)$ linearly independent solutions of GHE~(\ref{polar_GHE}) can be obtained from $(N+1)$ dimensional representation of $su(1,1)$. These solutions are of the form~(\ref{heun_polynomial}). Evidently, $\forall N \in N_0$, one obtains $(N+1)^2$ solutions of the form~(\ref{heun_polynomial}). The coefficients $k_i$ in the solutions and the eigenvalues $q_{\theta}$~(Eq.~(\ref{polar_GHE_params})) are given by the eigenvalues and the eigenvectors of the matrix~(\ref{GHE_tridiag_quasi}), which is the matrix form of the General Heun operator satisfying~(\ref{quasi_exact_condn}) for non-zero $\tau$ (here, $\tau^{(2)}_{\theta}$). These are \emph{additional} Heun polynomials, obtained as a result of casting the Heun equation in terms of generators~(\ref{newgenerators}). Putting these GHE solutions $y_{\pm}(z)$ back into Eq.~(\ref{polar_homotopy}), once again a countably infinite number of exact global solutions are obtained to the polar part~(\ref{Dirac_polar}) of the Dirac equation and the corresponding values of the separation constant ($\lambda$) are found using $q_{\theta}-\lambda$ relation given by Eq.~(\ref{polar_GHE_params}). Solutions upto triplet are listed in Table~\ref{tab_polar_sols_2}, among which, analytic expressions for the eigenvalues and eigenvectors have been provided for the singlet and the doublet and numerical expressions have been provided for the triplet for similar reasons as discussed for the $\left\{\sigma^{(1)}_{\theta},\tau^{(2)}_{\theta}\right\}$ case.
	
	\subsection{Behavior of the obtained polar solutions:}\label{subsec_polar_sol_nature}
	
	All the polar solutions listed in the Tables~\ref{tab_polar_sols_1} and \ref{tab_polar_sols_2} diverge at $z=0$ and $z=1$, i.e., at $\theta=\pi$ and $\theta=0$, respectively. The main point to note is that the polar solutions obtained exploiting the $su(1,1)$ symmetry of the polar GHE are divergent in the directions both along and opposite to the direction of acceleration of the black hole.
	
	{
	\renewcommand{\arraystretch}{1.6}
	\begin{table}[!ht]
		\caption{\label{tab_polar_sols_2}{First few solutions corresponding to $\left\{\sigma^{(2)}_{\theta},\tau^{(2)}_{\theta}\right\}$ pair for both polar functions $S_{\pm 1/2}$}}
		\centering
		\footnotesize
		\begin{tabular}{|@{}c@{}|c|@{}c@{}|@{}c@{}|}\hline
			\begin{tabular}{c}
				$su(1,1)$ \\ \textbf{Rep}
			\end{tabular} & $\eta $ & \textbf{$\lambda^2$} & \textbf{Solutions \big($S_{\pm 1/2}$\big)} \\
			\hline
			Singlet & $1/2$ & $5/8$ & $z^{-3/4}(z-1)^{-5/4}(z-3/2)^{3/2}$ \\
			\hline
			\multirow{4}{*}{Doublet} & \multirow{2}{*}{$1/4$} & $\frac{1}{16} \left(35+2\sqrt{166}\right)$ & $z^{5/4}(z-1)^{-17/12}(z-5/2)^{7/6}\left[\frac{1}{15} \left(-14+\sqrt{166}\right)z^{-3/2}+z^{-5/2}\right]$\\
			&  & $\frac{1}{16} \left(35-2\sqrt{166}\right)$ & $z^{5/4}(z-1)^{-17/12}(z-5/2)^{7/6}\left[\frac{1}{15} \left(-14-\sqrt{166}\right)z^{-3/2}+z^{-5/2}\right]$ \\ \cline{2-4}
			& \multirow{2}{*}{$3/4$} & $\frac{5}{16} \left(13+2\sqrt{22}\right)$ &  $z^{5/4}(z-1)^{-27/4}(z-7/6)^{13/2}\left[\frac{1}{21} \left(-26+5\sqrt{22}\right)z^{-3/2}+z^{-5/2}\right]$\\
			&  & $\frac{5}{16} \left(13-2\sqrt{22}\right)$ &  $z^{5/4}(z-1)^{-27/4}(z-7/6)^{13/2}\left[\frac{1}{21} \left(-26-5\sqrt{22}\right)z^{-3/2}+z^{-5/2}\right]$\\\hline
			\multirow{9}{*}{Triplet} & \multirow{3}{*}{$1/6$} & $8.8488$ & $z^{7/4}(z-1)^{-37/20}(z-7/2)^{11/10}\left[0.0035z^{-3/2}-0.1776z^{-5/2}+0.9841z^{-7/2}\right]$\\
			&  & $3.7158$ & $z^{7/4}(z-1)^{-37/20}(z-7/2)^{11/10}\left[-0.4430z^{-3/2}+0.8880z^{-5/2}-0.4577z^{-7/2}\right]$\\
			&  & $0.5605$ & $z^{7/4}(z-1)^{-37/20}(z-7/2)^{11/10}\left[0.6491z^{-3/2}-0.7222z^{-5/2}+0.2390z^{-7/2}\right]$\\ \cline{2-4}
			& \multirow{3}{*}{$1/2$} & $14.3609$ & $z^{7/4}(z-1)^{-17/4}(z-3/2)^{7/2}\left[0.0116z^{-3/2}-0.3194z^{-5/2}+0.9475z^{-7/2}\right]$\\
			&  & $6.4686$ & $z^{7/4}(z-1)^{-17/4}(z-3/2)^{7/2}\left[-0.0789z^{-3/2}+0.9225z^{-5/2}-0.3778z^{-7/2}\right]$\\
			&  & $1.0455$ & $z^{7/4}(z-1)^{-17/4}(z-3/2)^{7/2}\left[0.7551z^{-3/2}-0.6350z^{-5/2}+0.1633z^{-7/2}\right]$\\ \cline{2-4}
			& \multirow{3}{*}{$5/6$} & $20.0253$ & $z^{7/4}(z-1)^{-65/4}(z-11/10)^{31/2}\left[0.0163z^{-3/2}-0.3742z^{-5/2}+0.9272z^{-7/2}\right]$\\
			&  & $9.1147$ & $z^{7/4}(z-1)^{-65/4}(z-11/10)^{31/2}\left[-0.0948z^{-3/2}+0.9369z^{-5/2}-0.3365z^{-7/2}\right]$\\
			&  & $1.4850$ & $z^{7/4}(z-1)^{-65/4}(z-11/10)^{31/2}\left[0.7999z^{-3/2}-0.5855z^{-5/2}+0.1316z^{-7/2}\right]$\\ \hline
		\end{tabular}
	\end{table}
}
	
	\section{Heun polynomials solutions to the radial part:}\label{sec_radial}
	
	From Eqs.~(\ref{GHE_sigma_tau}) and (\ref{radial_GHE_params}), the $\sigma$ and $\tau$ values for the radial GHE~(\ref{radial_GHE}) are found to be
	\begin{equation}\label{radial_sigma_tau}
	\left\{\sigma^{(1)}_r,\sigma^{(2)}_r\right\}=\left\{0,-\dfrac{\eta^2-\eta\pm 4\iota M \omega}{4\eta(\eta-1)}\right\} \quad \mbox{and} \quad \left\{\tau^{(1)}_r,\tau^{(2)}_r\right\}=\left\{0,\dfrac{1}{4}\right\}.
	\end{equation}
	
	Among the four possible $\left\{\sigma_r,\tau_r\right\}$ pairs, it is easy to see that the pair $\big\{\sigma^{(1)}_r,\tau^{(2)}_r\big\}$ does not satisfy Heun polynomial condition~(\ref{quasi_exact_condn}). The pair $\big\{\sigma^{(1)}_r,\tau^{(1)}_r\big\}$ satisfies condition~(\ref{quasi_exact_condn}) for $N=0$. Hence, singlet solution of the form $z^{2\tau}P_N(z)$ (Eq.~(\ref{heun_polynomial})) is obtained to the GHE~(\ref{radial_GHE}), which is a constant in this case \big(since $\tau=\tau^{(1)}_r=0$ and $N=0$\big) and the corresponding eigenvalue $q_r=0$. Plugging this solution into Eq.~(\ref{radial_homotopy}) and using Eq.~(\ref{radial_Mobius}) the following solution is obtained to the radial part~(\ref{Dirac_radial}) of the Dirac equation (and the separation constant $\lambda$ is found using $q_r-\lambda$ relation given by Eq.~(\ref{radial_GHE_params})):
	\begin{equation}\label{constraint_free_radial_soln}
	R_{\pm 1/2}= \left\{\dfrac{Ar-\eta}{\eta\left(1-Ar\right)}\right\}^{\mp\dfrac{2\iota M \omega}{1-\eta^2}}
	\left\{\dfrac{\left(1-\eta\right)\left(1+Ar\right)}{2\eta\left(1-Ar\right)}\right\}^{\mp \dfrac{\iota M \omega}{\eta(1+\eta)}} \quad \mbox{for} \quad \lambda^2=0.
	\end{equation}
	For the other two pairs $\big\{\sigma^{(2)}_r,\tau^{(1)}_r\big\}$ and $\big\{\sigma^{(2)}_r,\tau^{(2)}_r\big\}$, one can look for the parametric conditions leading to the fulfilment of condition~(\ref{quasi_exact_condn}).
	
	\subsection{Conditions for Heun polynomials from the $\left\{\sigma^{(2)}_r,\tau^{(1)}_r\right\}$ pair}
	
	The demand that the pair $\big\{\sigma^{(2)}_r,\tau^{(1)}_r\big\}$ must satisfy condition~(\ref{quasi_exact_condn}), i.e., $\big(\sigma^{(2)}_r-\tau^{(1)}_r\big)=N/2$, where $N \in \mathbb{N}_0$, requires
	\begin{equation}
	\eta\left(\eta-1\right)\pm \dfrac{2\iota M \omega}{N+1/2}=0
	\end{equation}
	for $R_{\pm 1/2}$. For $R_{+1/2}$ this yields
	\begin{equation}\label{eta_Mw_reln_1_plus}
	\omega=\dfrac{\iota}{2M}\eta\left(\eta-1\right)\left(N+1/2\right) \quad \mbox{or} \quad \eta=\dfrac{1\pm \sqrt{1-\dfrac{8\iota M\omega}{N+1/2}}}{2}.
	\end{equation}
	Imposing the range~(\ref{etarange}) of $\eta$ one obtains for both expressions of $\eta$ given by Eq.~(\ref{eta_Mw_reln_1_plus})
	\begin{equation}
	\iota M\omega>0,
	\end{equation}
	which is satisfied by the expression~(\ref{eta_Mw_reln_1_plus}) for $\omega$. Similarly for $R_{-1/2}$, the analogue of Eq.~(\ref{eta_Mw_reln_1_plus}) is
	\begin{equation}\label{eta_Mw_reln_1_minus}
	\omega=-\dfrac{\iota}{2M}\eta\left(\eta-1\right)\left(N+1/2\right) \quad \mbox{or} \quad \eta=\dfrac{1\pm \sqrt{1+\dfrac{8\iota M\omega}{N+1/2}}}{2},
	\end{equation}
	which yields
	\begin{equation}
	\iota M \omega<0
	\end{equation}
	for both the expressions of $\eta$ given by Eq.~(\ref{eta_Mw_reln_1_minus}) and this is satisfied by the expression~(\ref{eta_Mw_reln_1_minus}) for $\omega$.  Thus, the allowed values of $\omega$ for the two radial functions $R_{\pm1/2}$ are exactly opposite in sign to each other. When these values are put back to the radial equations~(\ref{Dirac_radial}) (or the corresponding GHEs~(\ref{radial_GHE})), which differs only in the sign of $\omega$, the equations become identical. Consequently, the solutions to them become identical as well (modulo the interpretation). For the $\omega$ values given by Eqs.~(\ref{eta_Mw_reln_1_plus}) and (\ref{eta_Mw_reln_1_minus}) for $R_{+1/2}$ and $R_{-1/2}$ respectively, the finite $(N+1)$ dimensional representation space of $su(1,1)$ serves as the solutions space of the radial GHE~(\ref{radial_GHE}). Hence, $\forall N \in \mathbb{N}_0$, $(N+1)$ linearly independent Heun polynomial solutions of the form~(\ref{heun_polynomial}) are obtained where the coefficintes $k_i$ and the eigenvalues $q_r$ are given by the eigenvectors and eigenvalues of matrix~(\ref{GHE_tridiag_poly}) since $\tau=\tau^{(1)}_r=0$ in this case. Plugging these into Eq.~(\ref{radial_homotopy}) and using Eq.~(\ref{radial_Mobius}), (\ref{radial_expnts}) with $\omega$ values given by Eqs.~(\ref{eta_Mw_reln_1_plus}) and (\ref{eta_Mw_reln_1_minus}), one obtains, $\forall N \in \mathbb{N}_0$, the following expression for the radial functions $R_{\pm 1/2}$:
	\begin{equation}\label{generic_radial_soln_1}
	\left(\frac{Ar-\eta}{\eta\left(1-Ar\right)}\right)^{-\frac{(2N+1)\eta}{2(1+\eta)}}
		\left(\frac{\left(1-\eta\right)\left(1+Ar\right)}{2\eta\left(1-Ar\right)}\right)^{- \frac{(2N+1)(1-\eta)}{4(1+\eta)}}
		u^{2\tau^{(1)}_r}\sum_{i=0}^{N}k_iu^i.
	\end{equation}
	Solutions upto triplet are listed in Table~\ref{tab_radial_sols_1}. To get a better picture of the singularity structures of these solutions and to understand the behaviour at the Schwarzschild and Rindler horizons, it is instructive to simplify Eq.~(\ref{generic_radial_soln_1})
	\begin{eqnarray}\label{generic_radial_soln_1_simplified}
	\nonumber R_{\pm 1/2}= \left(\eta\left(1-Ar\right)\right)^{\frac{2N+1}{4}}
		\left(Ar-\eta\right)^{-\frac{(2N+1)\eta}{2(1+\eta)}}
		\left(\frac{\left(1-\eta\right)\left(1+Ar\right)}{2}\right)^{- \frac{(2N+1)(1-\eta)}{4(1+\eta)}}\\
	.\sum_{i=0}^{N}k_i\left(\frac{Ar(1-\eta)}{\eta(1-Ar)}\right)^i.
	\end{eqnarray}
	The behaviour at the Schwarzschild horizon is be governed by power of $(Ar-\eta)$, which appears only at the pre-factor multiplying the polynomial in solutions~(\ref{generic_radial_soln_1_simplified}). Now since $\eta(=2MA)$ is always a positive quantity, the exponent of $(Ar-\eta)$ in solution~(\ref{generic_radial_soln_1_simplified}) is always negative. Hence, near the Schwarzschild horizon $(r\to2M)$ the solution blows up like $(r-2M)^{-\frac{(2N+1)\eta}{2(1+\eta)}}$.
	
	The behaviour at the Rindler horizon $(r=1/A)$, on the other hand, depends on the power of $(1-Ar)$ in the pre-factor multiplying the polynomial and that in the polynomial. As stated above, the coefficients of the polynomial will be given by the eigenvectors of matrix~(\ref{GHE_tridiag_poly}). Now it may be noted that all entries in the first column of the matrix~(\ref{GHE_tridiag_poly}) are 0 except one which is given by $\alpha\beta$. If any one (or both) of $\alpha$ and $\beta$ is (are) zero, then the first column consists of $0$'s only. Hence, from matrix algebra, it is evident that a column vector with all $0$'s but the first entry (which is normalized to $1$) would be an eigenvector of that matrix. For the radial GHE, $\alpha_r=0$~(Eq.~(\ref{radial_GHE_params})), hence, $\forall N \in \mathbb{N}_0$, one among the $(N+1)$ eigenvectors will be an eigenvector of the abovesaid type. Therefore, $\forall N \in \mathbb{N}_0$, the polynomial with $k_0=1~(\neq 0)$ and $k_i=0,~i \in \{1,2,\dots,N\}$ represents a solution of the radial GHE, which is a constant. Evidently, for these solutions, the eigenvalue $q_r=\lambda^2/2\eta=0$, which means, the separation constant $\lambda=0$. The behaviour of these solutions at the Rindler horizon will be governed by the power of $(1-Ar)$ in the pre-factor only, which is a positive number $(2N+1)/4$, $\forall N \in \mathbb{N}_0$. Thus these solutions converge to $0$ at the Rindler horizon. In Table~\ref{tab_radial_sols_1}, these solutions are given by the first row for every $(N+1)$-plet. Now $\forall N(>0) \in \mathbb{N}$, there are $N$ other solutions for whom $k_N \neq 0$ and hence the maximum divergence coming from the polynomial will be $(1-Ar)^{-N}$. Combining with the pre-factor power $(2N+1)/4$, the maximum divergence is found to behave as $(1-2N)/4$, $\forall N(>0) \in \mathbb{N}$. Hence, $N$ among the $(N+1)$ linearly independent radial solutions forming each $(N+1)$-plet (for $N(>0) \in \mathbb{N}$), diverge as power $(1-2N)/4$. These are given in Table~\ref{tab_radial_sols_1} by second row onwards of each $(N+1)$-plet (except the singlet).
	
	{
	\renewcommand{\arraystretch}{1.8}
	\begin{sidewaystable}
		\caption{\label{tab_radial_sols_1}{First few solutions corresponding to $\left\{\sigma^{(2)}_r,\tau^{(1)}_r\right\}$ pair for both radial functions $R_{\pm 1/2}$ ($\omega_{\pm}$ denote the angular frequency for $R_{\pm 1/2}$)}}
		\centering
		\begin{tabular}{|c|c|c|c|c|}
			\hline
			\begin{tabular}{c}
				$N$ \\ \textbf{value}
			\end{tabular} & \begin{tabular}{c}
			$su(1,1)$ \\ \textbf{Rep}
		\end{tabular} & \textbf{$\omega_{\pm}$} & $\lambda^2$ & Solutions $R_{\pm 1/2}$ \\
			\hline
			$0$ & $1$(Singlet) & $\pm \frac{\iota}{4M}\scriptstyle \eta\left(\eta-1\right)$ & 0 & $\left(\frac{Ar-\eta}{\eta\left(1-Ar\right)}\right)^{-\frac{\eta}{2(1+\eta)}}
			\left(\frac{\left(1-\eta\right)\left(1+Ar\right)}{2\eta\left(1-Ar\right)}\right)^{- \frac{1-\eta}{4(1+\eta)}}$ \\
			\hline
			\multirow{2}{*}{$1$} & \multirow{2}{*}{$2$(Doublet)} & \multirow{2}{*}{$\pm \frac{3\iota}{4M}\scriptstyle \eta\left(\eta-1\right)$} & 0 & $\left(\frac{Ar-\eta}{\eta\left(1-Ar\right)}\right)^{-\frac{3\eta}{2(1+\eta)}}
				\left(\frac{\left(1-\eta\right)\left(1+Ar\right)}{2\eta\left(1-Ar\right)}\right)^{- \frac{3(1-\eta)}{4(1+\eta)}}$ \\
				\cline{4-5} 
			& & & $\scriptstyle 1-3\eta$ & $\left(\frac{Ar-\eta}{\eta\left(1-Ar\right)}\right)^{-\frac{3\eta}{2(1+\eta)}}
			\left(\frac{\left(1-\eta\right)\left(1+Ar\right)}{2\eta\left(1-Ar\right)}\right)^{- \frac{3(1-\eta)}{4(1+\eta)}}$ $\left\{\frac{1-\eta}{2\left(3\eta-1\right)}+\left(\frac{Ar(1-\eta)}{\eta\left(1-Ar\right)}\right)\right\}$\\
			\hline  
			\multirow{5}{*}{$2$} & \multirow{5}{*}{$3$(Triplet)} & \multirow{5}{*}{$\pm \frac{5\iota}{4M}\scriptstyle \eta\left(\eta-1\right)$} & 0 & $\left(\frac{Ar-\eta}{\eta\left(1-Ar\right)}\right)^{-\frac{5\eta}{2(1+\eta)}}
				\left(\frac{\left(1-\eta\right)\left(1+Ar\right)}{2\eta\left(1-Ar\right)}\right)^{- \frac{5(1-\eta)}{4(1+\eta)}}$ \\
				\cline{4-5}
			& & & \multirow{2}{*}{$\scriptstyle (5-15
				\eta-\sqrt{3} \sqrt{19 \eta^2-10 \eta+3})/2$} & \multicolumn{1}{|l|}{$\left(\frac{Ar-\eta}{\eta\left(1-Ar\right)}\right)^{-\frac{5\eta}{2(1+\eta)}}
				\left(\frac{\left(1-\eta\right)\left(1+Ar\right)}{2\eta\left(1-Ar\right)}\right)^{- \frac{5(1-\eta)}{4(1+\eta)}}$} \\
			& & & & $.\left\{\frac{(\eta-1) \left(9 \eta-3-\sqrt{3} \sqrt{19 \eta^2-10 \eta+3}\right)}{4 \eta \left(15
				\eta-5+\sqrt{3} \sqrt{19 \eta^2-10 \eta+3}\right)}-\frac{9 \eta-3-\sqrt{3} \sqrt{19 \eta^2-10 \eta+3}}{4 \eta}\left(\frac{Ar(1-\eta)}{\eta\left(1-Ar\right)}\right)+\left(\frac{Ar(1-\eta)}{\eta\left(1-Ar\right)}\right)^2\right\}$ \\
			\cline{4-5}
			& & & \multirow{2}{*}{$\scriptstyle (5-15
				\eta+\sqrt{3} \sqrt{19 \eta^2-10 \eta+3})/2$} & \multicolumn{1}{|l|}{$\left(\frac{Ar-\eta}{\eta\left(1-Ar\right)}\right)^{-\frac{5\eta}{2(1+\eta)}}
				\left(\frac{\left(1-\eta\right)\left(1+Ar\right)}{2\eta\left(1-Ar\right)}\right)^{- \frac{5(1-\eta)}{4(1+\eta)}}$} \\
			& & & & $.\left\{\frac{(\eta-1) \left(9 \eta-3+\sqrt{3} \sqrt{19 \eta^2-10 \eta+3}\right)}{4\eta \left(15
				\eta-5-\sqrt{3} \sqrt{19 \eta^2-10 \eta+3}\right)}-\frac{9 \eta-3+\sqrt{3} \sqrt{19 \eta^2-10 \eta+3}}{4 \eta}\left(\frac{Ar(1-\eta)}{\eta\left(1-Ar\right)}\right)+\left(\frac{Ar(1-\eta)}{\eta\left(1-Ar\right)}\right)^2\right\}$ \\
			\hline
		\end{tabular}
		
		\bigskip 
		\caption{\label{tab_radial_sols_2}{First few solutions corresponding to $\left\{\sigma^{(2)}_r,\tau^{(2)}_r\right\}$ pair for both radial functions $R_{\pm 1/2}$ ($\omega_{\pm}$ denote the angular frequency for $R_{\pm 1/2}$)}}
		\centering
		\begin{tabular}{|c|c|c|c|c|}
			\hline
			\begin{tabular}{c}
				$N$ \\ \textbf{value}
			\end{tabular} & \begin{tabular}{c}
			$su(1,1)$ \\ \textbf{Rep}
		\end{tabular} & \textbf{$\omega_{\pm}$} & $\lambda^2$ & Solutions $R_{\pm 1/2}$ \\
		\hline
		$0$ & $1$(Singlet) & $\pm \frac{\iota}{2M}\scriptstyle \eta\left(\eta-1\right)$ & $\scriptstyle (1-3\eta)/4$ & $\left(\frac{Ar-\eta}{\eta\left(1-Ar\right)}\right)^{-\frac{\eta}{(1+\eta)}}
		\left(\frac{\left(1-\eta\right)\left(1+Ar\right)}{2\eta\left(1-Ar\right)}\right)^{- \frac{1-\eta}{2(1+\eta)}}\left(\frac{Ar(1-\eta)}{\eta\left(1-Ar\right)}\right)^{1/2}$ \\
		\hline
		\multirow{2}{*}{$1$} & \multirow{2}{*}{$2$(Doublet)} & \multirow{2}{*}{$\pm \iota\scriptstyle \eta\left(\eta-1\right)$} & $\scriptstyle \left(5-15\eta-2\sqrt{2}\sqrt{2-9\eta+15\eta^2}\right)/4$ & $\left(\frac{Ar-\eta}{\eta\left(1-Ar\right)}\right)^{-\frac{2\eta}{(1+\eta)}}
		\left(\frac{\left(1-\eta\right)\left(1+Ar\right)}{2\eta\left(1-Ar\right)}\right)^{- \frac{1-\eta}{(1+\eta)}}
		\left(\frac{Ar(1-\eta)}{\eta\left(1-Ar\right)}\right)^{1/2}$ $\left\{\frac{2-6\eta+\sqrt{2}\sqrt{2-9\eta+15\eta^2}}{2\eta}+\left(\frac{Ar(1-\eta)}{\eta\left(1-Ar\right)}\right)\right\}$\\
		\cline{4-5} 
		& & & $\scriptstyle \left(5-15\eta+2\sqrt{2}\sqrt{2-9\eta+15\eta^2}\right)/4$ & $\left(\frac{Ar-\eta}{\eta\left(1-Ar\right)}\right)^{-\frac{2\eta}{(1+\eta)}}
		\left(\frac{\left(1-\eta\right)\left(1+Ar\right)}{2\eta\left(1-Ar\right)}\right)^{- \frac{1-\eta}{(1+\eta)}}
		\left(\frac{Ar(1-\eta)}{\eta\left(1-Ar\right)}\right)^{1/2}$ $\left\{\frac{2-6\eta-\sqrt{2}\sqrt{2-9\eta+15\eta^2}}{2\eta}+\left(\frac{Ar(1-\eta)}{\eta\left(1-Ar\right)}\right)\right\}$\\
		\hline
	\end{tabular}
	\end{sidewaystable}
}
	
	\subsection{Conditions for Heun polynomials from the $\left\{\sigma^{(2)}_r,\tau^{(2)}_r\right\}$ pair: additional solutions}
	
	The demand that the pair $\big\{\sigma^{(2)}_r,\tau^{(2)}_r\big\}$ must satisfy condition~(\ref{quasi_exact_condn}), i.e., $\big(\sigma^{(2)}_r-\tau^{(2)}_r\big)=N/2$, where $N \in \mathbb{N}_0$, requires
	\begin{equation}
	\eta\left(\eta-1\right)\pm \dfrac{2\iota M \omega}{N+1}=0
	\end{equation}
	for $R_{\pm 1/2}$. For $R_{+1/2}$ this yields
	\begin{equation}\label{eta_Mw_reln_2_plus}
	\omega=\dfrac{\iota}{2M}\eta\left(\eta-1\right)\left(N+1\right) \quad \mbox{or} \quad \eta=\dfrac{1\pm \sqrt{1-\dfrac{8\iota M\omega}{N+1}}}{2}.
	\end{equation}
	Imposing the range~(\ref{etarange}) of $\eta$ one obtains for both expressions of $\eta$ given by Eq.~(\ref{eta_Mw_reln_2_plus})
	\begin{equation}
	\iota M\omega>0,
	\end{equation}
	which is satisfied by the expression~(\ref{eta_Mw_reln_2_plus}) for $\omega$. Similarly for $R_{-1/2}$, the analogue of Eq.~(\ref{eta_Mw_reln_2_plus}) is
	\begin{equation}\label{eta_Mw_reln_2_minus}
	\omega=-\dfrac{\iota}{2M}\eta\left(\eta-1\right)\left(N+1\right) \quad \mbox{or} \quad \eta=\dfrac{1\pm \sqrt{1+\dfrac{8\iota M\omega}{N+1}}}{2},
	\end{equation}
	which yields
	\begin{equation}
	\iota M \omega<0
	\end{equation}
	for both the expressions of $\eta$ given by Eq.~(\ref{eta_Mw_reln_2_minus}) and this is satisfied by the expression~(\ref{eta_Mw_reln_2_minus}) for $\omega$. Once Again, the allowed values of $\omega$ for the two radial functions $R_{\pm1/2}$ are exactly opposite in sign to each other. When these values are put back to the radial equations~(\ref{Dirac_radial}) (or the corresponding GHEs~(\ref{radial_GHE})), which differs only in the sign of $\omega$, the equations become identical. Consequently, the solutions to them become identical as well (modulo the interpretation). For the $\omega$ values given by Eqs.~(\ref{eta_Mw_reln_2_plus}) and (\ref{eta_Mw_reln_2_minus}) for $R_{+1/2}$ and $R_{-1/2}$ respectively, the finite $(N+1)$ dimensional representation space of $su(1,1)$ serves as the solutions space of the radial GHE~(\ref{radial_GHE}). $(N+1)$ linearly independent Heun polynomial solutions of the form~(\ref{heun_polynomial}) are obtained where the coefficintes $k_i$ and the eigenvelues $q_r$ are given by the eigenvectors and eigenvalues of matrix~(\ref{GHE_tridiag_quasi}) since $\tau=\tau^{(2)}_r \neq 0$ in this case. These are \emph{additional} Heun polynomials, obtained as a result of casting the Heun equation in terms of generators~(\ref{newgenerators}).
	
	Plugging these into Eq.~(\ref{radial_homotopy}) and using Eq.~(\ref{radial_Mobius}), (\ref{radial_expnts}) with $\omega$ values given by Eqs.~(\ref{eta_Mw_reln_2_plus}) and (\ref{eta_Mw_reln_2_minus}), one obtains, $\forall N \in \mathbb{N}_0$, the following expression for the radial functions $R_{\pm 1/2}$:
	\begin{equation}\label{generic_radial_soln_2}
	\left(\frac{Ar-\eta}{\eta\left(1-Ar\right)}\right)^{-\frac{(N+1)\eta}{(1+\eta)}}
		\left(\frac{\left(1-\eta\right)\left(1+Ar\right)}{2\eta\left(1-Ar\right)}\right)^{- \frac{(N+1)(1-\eta)}{2(1+\eta)}}
		u^{2\tau^{(2)}_r}\sum_{i=0}^{N}k_iu^i.
	\end{equation}
	Solutions up to doublet are listed in Table~\ref{tab_radial_sols_2}. To have an idea of the singularity structures of these solutions and to understand their behaviour at the Schwarzschild and Rindler horizons, Eq.~(\ref{generic_radial_soln_2}) is simplified below.
	\begin{eqnarray}\label{generic_radial_soln_2_simplified}
	\nonumber R_{\pm 1/2}= \sqrt{Ar}\left(Ar-\eta\right)^{-\frac{(N+1)\eta}{(1+\eta)}}
		\left(\frac{\left(1-\eta\right)\left(1+Ar\right)}{2}\right)^{- \frac{(N+1)(1-\eta)}{2(1+\eta)}}\left(1-\eta\right)^{1/2}\\
	.\sum_{i=0}^{N}k_i\left(\frac{Ar(1-\eta)}{\eta(1-Ar)}\right)^i
	\end{eqnarray}
	The power of $(Ar-\eta)$, which appears only in the pre-factor multiplying the polynomial in Eq.~(\ref{generic_radial_soln_2_simplified}), determines the behaviour at the Schwarzschild horizon. The fact that $\eta=(2MA)$ is always positive makes the power of $(Ar-\eta)$ always negative, which implies that near Schwarzschild horizon the solution~(\ref{generic_radial_soln_2_simplified}) diverges like $(r-2M)^{-\frac{(N+1)\eta}{1+\eta}}$.
	
	The behaviour of solutions~(\ref{generic_radial_soln_2_simplified}) at the Rindler horizon~$(r=1/A)$ is governed by the power of $(1-Ar)$ in the solutions. This term does not appear at all in the pre-factor multiplying the polynomial. Hence, the singlet solution, for which the polynomial is a constant, is well-behaved at the Rindler horizon. For doublet$(N=1)$ onwards the solutions diverge as $(1-Ar)^{-N}$, as is evident from the form of the polynomial.
	
	\section{Conclusion}\label{sec_conclusion}
	
	The dynamics of massless Dirac particles in the $C$-metric has been studied. The radial and polar parts of the Dirac equation were known~\cite{bini_bit_ger} to be equivalent to the General Heun Equation up to some transformation. Exploiting an underlying $su(1,1)$ symmetry of the GHE~\cite{my_2nd_paper}, both the radial and the polar GHEs have been found to admit a pair of countably infinite sets of (quasi-)polynomial type global solutions. The first few solutions of each class have been listed. These solutions are expected to provide valuable insights into the behavior of massless spin-$1/2$ particles in the $C$-metric.
	
	\section*{Acknowledgements:} The author would like to thank Dr. Ritesh K. Singh for his valuable suggestions.
	
	\appendix
	
	\begin{landscape}
		\centering
		\Huge Appendix
		\normalsize
		\section{Matrix forms $\mathcal{H}^p_{GHE}$~\cite{ciftci_hall_saad_dogu}/$\mathcal{H}^q_{GHE}$~\cite{my_2nd_paper} of the GHE when it admits $(N+1)$ linearly independent polynomial/quasi-polynomial solutions}\label{appendix_matrix_forms}
		\begin{equation}\label{GHE_tridiag_poly}
		\mathcal{H}^p_{GHE}=
		\begin{bmatrix}[1.5]
		0 & \substack{a\gamma} & 0 & 0&\dots&0&0 \\
		
		\substack{\alpha\beta} & \substack{-(a(\delta+\gamma)+\epsilon+\gamma)} &  \substack{2a(1+\gamma)}& 0&\dots&0&0 \\
		
		0     & \substack{\alpha\beta+(\gamma+\epsilon+\delta)}     & \substack{-2(a+1)-2(a(\delta+\gamma)+\epsilon+\gamma))}&\substack{3(-2(a(\delta+\gamma)+\epsilon+\gamma)+a\gamma)}&\dots&0&0\\
		
		\vdots&\vdots&\vdots&\vdots&\vdots&\vdots&\vdots\\ 
		
		0&0&0&0&\dots&\substack{\alpha\beta+(N-1)[(N-2)\\-(\gamma+\epsilon+\delta)]}&\substack{-N(N-1)(1+a)\\-N(a(\delta+\gamma)+\epsilon+\delta)}
		\end{bmatrix}
		\end{equation}
		
		\vspace{2cm}
		
		\begin{equation}\label{GHE_tridiag_quasi}
		\mathcal{H}^{q}_{GHE}=
		\begin{bmatrix}[1.8]
		\substack{(N-0)(1+0)-(N-0)\gamma\\-\gamma\delta +\beta\gamma+a\gamma\delta-\beta-\delta(a-1)}
		&\substack{2a-a\gamma}&0&0&\dots&0&0&0\\
		
		\substack{N(\gamma-\beta)-N.1}&\substack{(N-1)(1+1)-(N-1)\gamma\\-\gamma\delta +\beta\gamma+a\gamma\delta-2\beta\\-2\delta(a-1)+a\gamma-2a}&\substack{6a-2a\gamma}&0&\dots&0&0&0\\
		
		0&\substack{(N-1)(\gamma-\beta)-(N-1).2}&\substack{(N-2)(1+2)-(N-2)\gamma\\-\gamma\delta +\beta\gamma+a\gamma\delta-3\beta\\-3\delta(a-1)+2a\gamma-6a}&\substack{12a-3a\gamma}&\dots&0&0&0\\
		
		\vdots&\vdots&\vdots&\vdots&\vdots&\vdots&\vdots&\vdots\\
		
		0&0&0&0&\dots&\substack{2(\gamma-\beta)\\-2.(N-1)}&\substack{1.N-\gamma-\gamma\delta +\beta\gamma\\+a\gamma\delta-N\beta-N\delta(a-1)\\+(N-1)a\gamma-N(N-1)a}&\substack{(N+1)(N+2)a\\-(N+1)a\gamma}\\
		
		0&0&0&0&\dots&0&\substack{(\gamma-\beta)-1.N}&\substack{-\gamma\delta+\beta\gamma+a\gamma\delta\\-(N+1)\beta-(N+1)\delta(a-1)\\+Na\gamma-N(N+1)a}
		\end{bmatrix}
		\end{equation}
	\end{landscape}

\end{document}